\documentclass[american,nofootinbib]{revtex4}
\usepackage{graphicx}
\usepackage{amssymb}

\makeatletter

\usepackage{amsthm}

\theoremstyle{plain}
\theoremstyle{definition}
\newtheorem{defn}{Definition}
\theoremstyle{remark}
\newtheorem*{rem*}{Remark}
\theoremstyle{remark}

\theoremstyle{plain}    
\newtheorem{lem}{Lemma} 
\theoremstyle{plain}    
\newtheorem{thm}{Theorem} 
\theoremstyle{plain}    
\newtheorem{cor}{Corollary} 
\theoremstyle{remark}    
\newtheorem*{acknowledgement*}{Acknowledgement} 
\theoremstyle{plain}
\newtheorem{exm}{Example}

\usepackage{babel}
\makeatother
\begin{document}

\title{The Generalized Lyapunov Theorem and its Application to Quantum Channels}

\author{Daniel Burgarth$^{1}$}

\author{Vittorio Giovannetti$^{2}$}

\affiliation{$^{1}$Department of Physics \& Astronomy, University College London,
Gower Street, London WC1E 6BT, UK \\
 $^{2}$NEST CNR-INFM \& Scuola Normale Superiore, Piazza dei Cavalieri
7, I-56126 Pisa, Italy}

\begin{abstract}
We give a simple and physically intuitive necessary and sufficient
condition for a map acting on a compact metric space to be mixing
(i.e. infinitely many applications of the map transfer any input into
a fixed convergency point). This is a generalization of the {}``Lyapunov
direct method''. First we prove this theorem in topological spaces
and for arbitrary continuous maps. Finally we apply our theorem to
maps which are relevant in Open Quantum Systems and Quantum Information,
namely Quantum Channels. In this context we also discuss the relations
between mixing and ergodicity (i.e. the property that there exist
only a single input state which is left invariant by a single application
of the map) showing that the two are equivalent when the invariant
point of the ergodic map is pure. 
\end{abstract}
\maketitle

\section{Introduction}

Repetitive applications of the same transformation is the key ingredient
of many controls techniques. In quantum information processing~\cite{NIELSEN}
they have been exploited to inhibit the decoherence of a system by
frequently perturbing its dynamical evolution~\cite{VIOLA1,VIOLA2,VIOLA3,VITALI,SIMON}
(\emph{Bang-Bang control}) or to improve the fidelity of quantum gates~\cite{FRANSON}
by means of frequent measurements (\textit{quantum Zeno-effec}t~\cite{PERES}).
Recently analogous strategies have also been proposed in the context
of state preparation~\cite{KUMMERER,WELLENS,HOMOGENIZATION1,HOMOGENIZATION2,TERHAL,YUASA1,YUASA2}
and quantum communication~\cite{MEMORYSWAP,DUALRAIL,RANDOMRAIL,MULTIRAIL}.
In Refs.~\cite{HOMOGENIZATION1,HOMOGENIZATION2} for instance, a
\emph{homogenization} protocol was presented which allows one to transform
any input state of a qubit into a some pre-fixed target state by repetitively
coupling it with an external bath. A similar \emph{thermalization}
protocol was discussed in Ref.~\cite{TERHAL} to study the efficiency
of simulating classical equilibration processes on a quantum computer.
In Refs.~\cite{YUASA1,YUASA2} repetitive interactions with an externally
monitored environment were instead exploited to implement {\em purification}
schemes which would allow one to extract pure state components from
arbitrary mixed inputs. An application to quantum communication of
similar strategies has been finally given in Refs.~\cite{MEMORYSWAP,DUALRAIL,RANDOMRAIL,MULTIRAIL}
where sequences of repetitive operations were used to boost the efficiency
of quantum information transmission along spin chains.

\begin{figure}
\includegraphics[width=6cm]{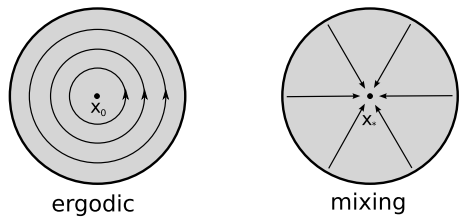}

\caption{\label{F1}Schematic examples of the orbits of a ergodic and a mixing
map.}
\end{figure}

The common trait of the proposals~\cite{KUMMERER,WELLENS,HOMOGENIZATION1,HOMOGENIZATION2,TERHAL,YUASA1,YUASA2,MEMORYSWAP,DUALRAIL,RANDOMRAIL,MULTIRAIL}
is the requirement that repeated applications of a properly chosen
quantum operation $\tau$ converges to a fixed density matrix $x_{*}$
independently from the input state $x$ of the system, i.e. \begin{eqnarray}
\tau^{n}(x)\equiv\underbrace{\tau\circ\tau\circ\cdots\circ\tau}_{n}\;(x)\Big|_{n\rightarrow\infty}\longrightarrow\;\; x_{*}\;,\label{mixing0}\end{eqnarray}
 with {}``$\circ$'' representing the composition of maps. Following
the notation of Refs.~\cite{RAGINSKY,RICHTER} we call Eq.~(\ref{mixing0})
the {\em mixing} property of $\tau$. It is related with another
important property of maps, namely {\em ergodicity} (see Fig. \ref{F1}).
The latter requires the existence of a unique input state $x_{0}$
which is left invariant under a single application of the map%
\footnote{Definition~(\ref{ergodic0}) may sound unusual for readers who are
familiar with a definition of ergodicity from statistical mechanics,
where a map is called ergodic if its invariant sets have measure 0
or 1. The notion of ergodicity used in the case of a discrete time
evolution of a quantum system is different \cite{STRICTCONTRATIONS,RAGINSKY}.
Here, the map $\tau$ is not acting on a measurable space but on the
compact convex set of quantum states. A perhaps more intuitive and
equivalent definition of ergodicity based on the time average of observables
is given by Lemma~\ref{thm:ergodic} of the Appendix.%
} , i.e., \begin{eqnarray}
\tau(x)=x\qquad\Longleftrightarrow\qquad x=x_{0}\;.\label{ergodic0}\end{eqnarray}

Ergodicity and the mixing property are of high interest not only in
the context of the above quantum information schemes. They also occur
on a more fundamental level in statistical mechanics \cite{STREATER}
and open quantum systems \cite{OPENQUANTUM,ALICKI}, where one would
like to study irreversibility and relaxation to thermal equilibrium.

In the case of quantum transformations one can show that mixing maps
with convergency point $x_{*}$ are also ergodic with fixed point
$x_{0}=x_{*}$. The opposite implication however is not generally
true since there are examples of ergodic quantum maps which are not
mixing (see the following). Sufficient conditions for mixing have
been discussed both in the specific case of quantum channel~\cite{TERHAL,RAGINSKY,STRICTCONTRATIONS}
and in the more abstract case of maps operating on topological spaces~\cite{STREATER}.
In particular the Lyapunov direct method~\cite{STREATER} allows
one to prove that an ergodic map $\tau$ is mixing if there exists
a continuous functional $S$ which, for all points but the fixed one,
is strictly increasing under $\tau$. Here we strengthen this criterion
by weakening the requirement on $S$: our {\em generalized} Lyapunov
functions are requested only to have limiting values $S(\tau^{n}(x))|_{n\rightarrow\infty}$
which differ from $S(x)$ for all $x\neq x_{0}$. It turns out that
the existence of such $S$ is not just a {\em sufficient} condition
but also a {\em necessary} condition for mixing. Exploiting this
fact one can easily generalize a previous result on {\em strictly
contractive} maps~\cite{RAGINSKY} by showing that maps which are
{\em asymptotic deformations} (see Definition \ref{asymptdef})
are mixing. This has, unlike contractivity, the advantage of being
a property independent of the choice of metric (see however \cite{RICHTER}
for methods of finding {}``tight'' norms). In some cases, the generalized
Lyapunov method permits also to derive an optimal mixing condition
for quantum channels based on the quantum relative entropy. Finally
a slightly modified version of our approach which employs {\em multi-central}
Lyapunov functions yields a characterization of (not necessarily mixing)
maps which in the limit of infinitely many applications move all points
towards a proper \emph{subset} (rather than a single point) of the
input space.

The introduction of a generalized Lyapunov method seems to be sound
not only from a mathematical point of view, but also from a physical
point of view. In effect, it often happens that the informations available
on the dynamics of a system are only those related on its asymptotic
behavior (e.g. its thermalization process), its finite time evolution
being instead difficult to characterize. Since our method is explicitly
constructed to exploit asymptotic features of the mapping, it provides
a more effective way to probe the mixing property of the process.

Presenting our results we will not restrict ourself to the case of
quantum operations. Instead, following~\cite{STREATER} we will derive
them in the more general context of continuous maps operating on topological
spaces~\cite{TOPOLOGYBOOK}. This approach makes our results stronger
by allowing us to invoke only those hypothesis which, to our knowledge,
are strictly necessary for the derivation. It is important to stress
however that, as a particular instance, all the Theorems and Lemmas
presented in the paper hold for any linear, completely positive, trace
preserving map (i.e. quantum channel) operating on a compact subset
of normed vectors (i.e. the space of the density matrices of a finite
dimensional quantum system). Therefore readers who are not familiar
with topological spaces can simply interpret our derivations as if
they were just obtained for quantum channels acting on a finite dimensional
quantum system.

The paper is organized as follows. In Sec.~\ref{GLT} the generalized
Lyapunov method along with some minor results are presented in the
context of topological and metric spaces. Then quantum channels are
analyzed in Sec.~\ref{QC} providing a comprehensive summary of the
necessary and sufficient conditions for the mixing property of these
maps. Conclusions and remarks end the paper in Sec.~\ref{CONCLUSION}.

\section{Generalized Lyapunov Theorem}

\label{GLT}

\subsection{Topological spaces}

\begin{figure}[htbp]

\begin{centering}\includegraphics[width=0.4\columnwidth]{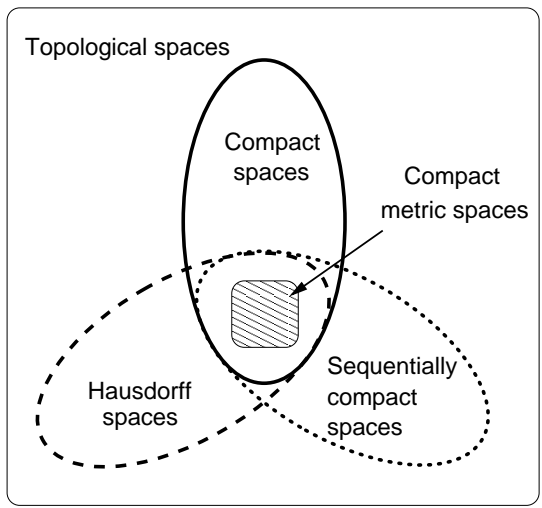}\par\end{centering}

\caption{\label{fig:topologic} Relations between topological spaces~\cite{TOPOLOGYBOOK}.
{\em Hausdorff topological spaces} have the property that any two
distinct points of the space can be separated by open neighborhoods:
for these sets any convergent sequence converges to a \emph{unique}
point of the set. {\em Compact topological spaces} are such that
any open cover of the set has a finite sub-cover. {\em Sequentially
compact topological spaces} are those for which the Bolzano-Weierstrass
theorem holds, i.e. every sequence has a convergent subsequence. Any
{\em compact metric spaces} is Hausdorff, compact, and sequentially
compact. The space of density matrices on which quantum channels are
defined, is a compact and convex subset of a normed vectors space
(the space of linear operators of the system) which, in the above
graphical representation fits within the set of compact metric spaces.}
\end{figure}

In this section we introduce the notation and derive our main result
(the Generalized Lyapunov Theorem). The properties of Hausdorff, compact
and sequentially compact topological spaces will be used~\cite{TOPOLOGYBOOK}.
For the sake of readability their definitions and their relations
are given in the caption of Fig.~\ref{fig:topologic}.

\begin{defn} Let $\mathcal{X}$ be a topological space and let $\tau:\mathcal{X}\rightarrow\mathcal{X}$
be a map. The sequence $x_{n}\equiv\tau^{n}(x)$, where $\tau^{n}$
is a short-hand notation for the $n-$fold composition of $\tau,$
is called the \emph{orbit} of $x.$ An element $x_{*}\in\mathcal{X}$
is called a \emph{fixed point} of $\tau$ if and only if \begin{eqnarray}
\tau(x_{*})=x_{*}\;.\label{defmixing}\end{eqnarray}
 $\tau$ is called \emph{ergodic} if and only if it has exactly one
fixed point. $\tau$ is called \emph{mixing} if and only if there
exists a \emph{convergence} point $x_{*}\in\mathcal{X}$ such that
any orbit converges to it, i.e. \begin{eqnarray}
\lim_{n\rightarrow\infty}x_{n}=x_{*}\quad\forall x\in\mathcal{X}\;.\label{defergo10}\end{eqnarray}
 \end{defn} \begin{rem*} Here we use the usual topological definition
of convergence, i.e. $\lim_{n\rightarrow\infty}x_{n}=x_{*}$ if and
only if for each open neighborhood $O(x_{*})$ of $x_{*}$ only finitely
many points of the sequence are not in $O(x_{*}).$ This clearly depends
on the topology, and there may exist many different points to which
a sequence converges to. For example, in the \emph{trivial topology}
of $\mathcal{X}$ where the only open sets are $\mathcal{X}$ and
the empty set, \emph{any} sequence is convergent to \emph{any} point.
On the other hand the uniqueness of the convergence point can be enforced
by requiring the topological set $\mathcal{X}$ to be Hausdorff~\cite{TOPOLOGYBOOK}
(see Fig.~\ref{fig:topologic} for an explicit definition of this
property). \end{rem*} A direct connection between ergodicity and
mixing can be established as follows. \begin{lem}\label{lem:hausdorff}
Let $\tau:\mathcal{X}\rightarrow\mathcal{X}$ be a continuous mixing
map on a topological Hausdorff space $\mathcal{X}.$ Then $\tau$
is ergodic. \end{lem} 

\begin{proof} Let $x_{*}$ be the convergence point of $\tau$ and
let $x\in\mathcal{X}$ arbitrary. Since $\tau$ is continuous we can
perform the limit in the argument of $\tau,$ i.e.\[
\tau(x_{*})=\tau\left(\lim_{n\rightarrow\infty}\tau^{n}(x)\right)=\lim_{n\rightarrow\infty}\tau^{n+1}(x)=x_{*},\]
 which shows that $x_{*}$ is a fixed point of $\tau$. To prove that
it is unique assume by contradiction that $\tau$ possesses a second
fixed point $y_{*}\neq x_{*}$. Then $\lim_{n\rightarrow\infty}\tau^{n}(y_{*})=y_{*}\neq x_{*}$,
so $\tau$ could not be mixing (since the limit is unique in a Hausdorff
space -- see Fig.~\ref{fig:topologic}). Hence $\tau$ is ergodic.
\end{proof} \begin{rem*} The converse is not true in general, i.e.
not every ergodic map is mixing (not even in Hausdorff topological
spaces). A simple counterexample is given by $\tau:[-1,1]\rightarrow[-1,1]$
with $\tau(x)\equiv-x$ and the usual topology of $\mathbb{R}$, which
is ergodic with fixed point $0,$ but not mixing since for $x\neq0$,
$\tau^{n}(x)=(-1)^{n}x$ is alternating between two points. A similar
counterexample will be discussed in the quantum channel section (see
Example~\ref{exm:ergodic}). \end{rem*} A well known criterion for
mixing is the existence of a \emph{Lyapunov function}~\cite{STREATER}.

\begin{defn} Let $\tau:\mathcal{X}\rightarrow\mathcal{X}$ be a map
on a topological space $\mathcal{X}.$ A continuous map $S:\mathcal{X}\rightarrow\mathbb{R}$
is called a \emph{(strict) Lyapunov function for $\tau$ around $x_{*}\in\mathcal{X}$}
if and only if\[
S\left(\tau(x)\right)>S(x)\quad\forall x\neq x_{*}.\]

\end{defn} \begin{rem*} At this point is is neither assumed that
$x_{*}$ \emph{is} a fixed point, nor that $\tau$ is ergodic. Both
follows from the theorem below. \end{rem*} \begin{thm}[Lyapunov function]
Let $\tau:\mathcal{X}\rightarrow\mathcal{X}$ be a continuous map
on a sequentially compact topological space $\mathcal{X}$. Let $S:\mathcal{X}\rightarrow\mathbb{R}$
be a Lyapunov function for $\tau$ around $x_{*}.$ Then $\tau$ is
mixing with the fixed point $x_{*}$. \end{thm} The proof of this
theorem is given in \cite{STREATER}. We will not reproduce it here,
because we will provide a general theorem that includes this as a
special case. In fact, we will show that the requirement of the strict
monotonicity can be \emph{much} weakened, which motivates the following
definition.

\begin{defn} Let $\tau:\mathcal{X}\rightarrow\mathcal{X}$ be a map
on a topological space $\mathcal{X}.$ A continuous map $S:\mathcal{X}\rightarrow\mathbb{R}$
is called a \emph{generalized Lyapunov function for $\tau$ around
$x_{*}\in\mathcal{X}$} if and only if the sequence $S\left(\tau^{n}(x)\right)$
is point-wise convergent for any $x\in\mathcal{X}$ and $S$ fulfills
\begin{equation}
S_{*}(x)\equiv\lim_{n\rightarrow\infty}S\left(\tau^{n}(x)\right)\neq S(x)\quad\forall x\neq x_{*}.\label{eq:weakinequality}\end{equation}

\end{defn} In general it may be difficult to prove the point-wise
convergence. However if $S$ is monotonic under the action of $\tau$
and the space is compact, the situation becomes considerably simpler.
This is summarized in the following Lemma.

\begin{lem} \label{lem:monotonlyapov}Let $\tau:\mathcal{X}\rightarrow\mathcal{X}$
be map on a compact topological space. A continuous map $S:\mathcal{X}\rightarrow\mathbb{R}$
which fulfills\begin{equation}
S\left(\tau(x)\right)\geqslant S(x)\quad\forall x\in\mathcal{X},\label{eq:monotonic}\end{equation}
 and\begin{equation}
S_{*}(x)\equiv\lim_{n\rightarrow\infty}S\left(\tau^{n}(x)\right)>S(x)\quad\forall x\neq x_{*}.\end{equation}
 for some fixed $x_{*}\in\mathcal{X}$ is a generalized Lyapunov function
for $\tau$ around $x_{*}$. \end{lem} \begin{proof} It only remains
to show the (point-wise) convergence of $S\left(\tau^{n}(x)\right)$.
Since $S$ is a continuous function on a compact space, it is bounded.
By Eq. (\ref{eq:monotonic}) the sequence is monotonic. Any bounded
monotonic sequence converges. \end{proof} \begin{cor} Let $\tau:\mathcal{X}\rightarrow\mathcal{X}$
be a map on a compact topological space. A continuous map $S:\mathcal{X}\rightarrow\mathbb{R}$
which fulfills\[
S\left(\tau(x)\right)\geqslant S(x)\quad\forall x\in\mathcal{X},\]
 and\[
S\left(\tau^{N}(x)\right)>S(x)\quad\forall x\neq x_{*},\]
 for some fixed $N\in\mathbb{N}$ and for some $x_{*}\in\mathcal{X}$
is a generalized Lyapunov function for $\tau$ around $x_{*}$. \end{cor}
\begin{rem*} This implies that a strict Lyapunov function is a generalized
Lyapunov function (with $N=1$). \end{rem*} We can now state the
main result of this section:

\begin{thm}[Generalized Lyapunov function] \label{thm:weaklyapov}
Let $\tau:\mathcal{X}\rightarrow\mathcal{X}$ be a continuous map
on a sequentially compact topological space $\mathcal{X}.$ Let $S:\mathcal{X}\rightarrow\mathbb{R}$
be a generalized Lyapunov function for $\tau$ around $x_{*}.$ Then
$\tau$ is mixing with fixed point $x_{*}$. \end{thm} \begin{proof}
Consider the orbit $x_{n}\equiv\tau^{n}(x)$ of a given $x\in\mathcal{X}.$
Because $\mathcal{X}$ is sequentially compact, the sequence $x_{n}$
has a convergent subsequence (see Fig.~\ref{fig:topologic}), i.e.
$\lim_{k\rightarrow\infty}x_{n_{k}}\equiv\tilde{x}$. Let us assume
that $\tilde{x}\neq x_{*}$ and show that this leads to a contradiction.
By Eq. (\ref{eq:weakinequality}) we know that there exists a finite
$N\in\mathbb{N}$ such that \begin{equation}
S\left(\tau^{N}(\tilde{x})\right)\neq S(\tilde{x}).\label{eq:contra}\end{equation}
 Since $\tau^{N}$ is continuous we can perform the limit in the argument,
i.e. $\lim_{k\rightarrow\infty}\tau^{N}\left(x_{n_{k}}\right)=\tau^{N}(\tilde{x})$.
Likewise, by continuity of $S$ we have \begin{equation}
\lim_{k\rightarrow\infty}S\left(x_{n_{k}}\right)=S(\tilde{x}),\label{eq:ss1}\end{equation}
 and on the other hand\begin{equation}
\lim_{k\rightarrow\infty}S\left(x_{N+n_{k}}\right)=\lim_{k\rightarrow\infty}S\left(\tau^{N}\left(x_{n_{k}}\right)\right)=S(\tau^{N}\tilde{x}),\label{eq:ss2}\end{equation}
 where the second equality stems from the continuity of the map $S$
and $\tau^{N}$. Because $S$ is a generalized Lyapunov function,
the sequence $S\left(x_{n}\right)$ is convergent. Therefore the subsequences
(\ref{eq:ss1}) and (\ref{eq:ss2}) must have the same limit. We conclude
that $S(\tau^{N}\tilde{x})=S(\tilde{x})$ which contradicts Eq. (\ref{eq:contra}).
Hence $\tilde{x}=x_{*}.$ Since we have shown that any convergent
subsequence of $\tau^{n}(x)$ converges to the same limit $x_{*}$,
it follows by Lemma~\ref{lem:subsequences} of the Appendix that
$\tau^{n}(x)$ is converging to $x_{*}.$ Since that holds for arbitrary
$x$, it follows that $\tau$ is mixing. \end{proof} There is an
even more general way of defining Lyapunov functions which we state
here for completeness. It requires the concept of the quotient topology
\cite{TOPOLOGYBOOK}.

\begin{defn} Let $\tau:\mathcal{X}\rightarrow\mathcal{X}$ be a map
on a topological space $\mathcal{X}.$ A continuous map $S:\mathcal{X}\rightarrow\mathbb{R}$
is called a \emph{multi-central Lyapunov function for $\tau$ around
$\mathcal{F}\subseteq\mathcal{X}$} if and only if the sequence $S\left(\tau^{n}(x)\right)$
is point-wise convergent for any $x\in\mathcal{X}$ and if $S$ and
$\tau$ fulfill the following three conditions: $S$ is constant on
$\mathcal{F}$, $\tau(\mathcal{F})\subseteq\mathcal{F}$, and \[
S_{*}(x)\equiv\lim_{n\rightarrow\infty}S\left(\tau^{n}(x)\right)\neq S(x)\quad\forall x\notin\mathcal{F}.\]
 \end{defn} For these functions we cannot hope that the orbit is
mixing. We can however show that the orbit is {}``converging'' to
the set $\mathcal{F}$ in the following sense:

\begin{thm}[Multi-central Lyapunov function] Let $\tau:\mathcal{X}\rightarrow\mathcal{X}$
be a continuous map on a sequentially compact topological space $\mathcal{X}.$
Let $S:\mathcal{X}\rightarrow\mathbb{R}$ be a multi-central Lyapunov
function for $\tau$ around $\mathcal{F}.$ Let $\varphi:\mathcal{X}\rightarrow\mathcal{X}/\mathcal{F}$
be the continuous mapping into the quotient space (i.e. $\varphi(x)=[x]$
for $x\in\mathcal{X}\backslash\mathcal{F}$ and $\varphi(x)=[\mathcal{F}]$
for $x\in\mathcal{F})$. Then $\tilde{\tau}:\mathcal{X}/\mathcal{F}\rightarrow\mathcal{X}/\mathcal{F}$
given by $\tilde{\tau}([x])=\varphi\left(\tau\left(\varphi^{-1}([x])\right)\right)$
is mixing with fixed point $[\mathcal{F}]$. \end{thm} \begin{proof}
First note that $\tilde{\tau}$ is well defined because $\varphi$
is invertible on $\mathcal{X}/\mathcal{F}\backslash[\mathcal{F}]$
and $\tau(\mathcal{F})\subseteq\mathcal{F},$ so that $\tilde{\tau}([\mathcal{F}])=[\mathcal{F}]$.
Since $\mathcal{X}$ is sequentially compact, the quotient space $\mathcal{X}/\mathcal{F}$
is also sequentially compact. Note that for $O$ open, $\tilde{\tau}^{-1}(O)=\varphi\left(\tau^{-1}\left(\varphi^{-1}\left(O\right)\right)\right)$
is the image of $\varphi$ of an open set in $\mathcal{X}$ and therefore
(by definition of the quotient topology) open in $\mathcal{X}/\mathcal{F}.$
Hence $\tilde{\tau}$ is continuous. The function $\tilde{S}([x]):\mathcal{X}/\mathcal{F}\rightarrow\mathcal{X}/\mathcal{F}$
given by $\tilde{S}([x])=S(\varphi^{-1}([x]))$ is continuous and
easily seen to be a generalized Lyapunov function around $[\mathcal{F}].$
By Theorem \ref{thm:weaklyapov} it follows that $\tilde{\tau}$ is
mixing. \end{proof}

\subsection{Metric spaces}

We now show that for the particular class of compact topological sets
which posses a metric, the existence of a generalized Lyapunov function
is also a necessary condition for mixing. In this context the convergence
of a sequence is defined with respect to the distance function $d(\cdot,\cdot):\mathcal{X}\times\mathcal{X}\rightarrow\mathbb{R}$
on the space, so that for instance Eq.~(\ref{defmixing}) requires
$\lim_{n\rightarrow\infty}d(x_{n},x_{*})=0$.

\begin{thm}[Lyapunov criterion] \label{thm:glyap} Let $\tau:\mathcal{X}\rightarrow\mathcal{X}$
be a continuous map on a compact metric space $\mathcal{X}.$ Then
$\tau$ is mixing with fixed point $x_{*}$ if and only if a generalized
Lyapunov function around $x_{*}$ exists. \end{thm} \begin{proof}
Firstly, in metric spaces compactness and sequential compactness are
equivalent, so the requirements of Theorem \ref{thm:weaklyapov} are
met. Secondly, for any mixing map $\tau$ with fixed point $x_{*},$
a generalized Lyapunov function around $x_{*}$ is given by $S(x)\equiv d(x_{*},x)$.
In fact, it is continuous because of the continuity of the metric
and satisfies \[
\lim_{n\rightarrow\infty}S\left(\tau^{n}(x)\right)=d(x_{*},x_{*})=0\leqslant d(x_{*},x)=S(x),\]
 where the equality holds if and only $x=x_{*}.$ We call $d(x_{*},x)$
the \emph{trivial generalized Lyapunov function}. \end{proof} \begin{rem*}
In the above Theorem we have not used all the properties of the metric.
In fact a continuous \emph{semi-metric} (i.e. without the triangle
inequality) would suffice.

The trivial Lyapunov function requires knowledge of the fixed point
of the map. There is another way of characterizing mixing maps as
those which bring elements closer to \emph{each other} (rather than
closer to the fixed point). \end{rem*} \begin{defn} A map $\tau:\mathcal{X}\rightarrow\mathcal{X}$
is on a metric space is called a \emph{non-expansive map} if and only
if \[
d(\tau(x),\tau(y))\leqslant d(x,y)\quad\forall x,y\in\mathcal{X},\]
 a \emph{weak contraction} if and only if \[
d(\tau(x),\tau(y))<d(x,y)\quad\forall x,y\in\mathcal{X},\, x\neq y,\]
 and a \emph{strict contraction} if and only if there exists a $k<1$
such that\[
d(\tau(x),\tau(y))\leqslant k\, d(x,y)\quad\forall x,y\in\mathcal{X}\,.\]
 \end{defn} \begin{rem*} The notation adopted here is slightly different
from the definitions adopted by other Authors~\cite{RAGINSKY,RUSKAI,WERNER}
who use contraction to indicate our non-expansive maps. Our choice
is motivated by the need to clearly distinguish between non-expansive
transformation and weak contractions. \end{rem*}

We can generalize the above definition in the following way:

\begin{defn} \label{asymptdef}A map $\tau:\mathcal{X}\rightarrow\mathcal{X}$
on a metric space is called an \emph{asymptotic deformation} if and
only if the sequence $d(\tau^{n}(x),\tau^{n}(y))$ converges point-wise
for all $x,y\in\mathcal{X}$ and\[
\lim_{n\rightarrow\infty}d(\tau^{n}(x),\tau^{n}(y))\neq d(x,y)\quad\forall x,y\in\mathcal{X},\, x\neq y.\]

\end{defn} \begin{rem*} Let $\tau:\mathcal{X}\rightarrow\mathcal{X}$
be a {non-expansive map} on a metric space $\mathcal{X},$ and let
\[
\lim_{n\rightarrow\infty}d(\tau^{n}(x),\tau^{n}(y))<d(x,y)\quad\forall x,y\in\mathcal{X},\, x\neq y.\]
 Then $\tau$ is an asymptotic deformation. Any weak contraction is
an asymptotic deformation. \end{rem*} \begin{thm}[Asymptotic deformations]
\label{thm:weakBanach} Let $\tau:\mathcal{X}\rightarrow\mathcal{X}$
be a continuous map on a compact metric space $\mathcal{X}$ with
at least one fixed point. Then $\tau$ is mixing if and only if $\tau$
is an asymptotic deformation. \end{thm} \begin{proof} Firstly assume
that $\tau$ is an asymptotic deformation. Let $x_{*}$ be a fixed
point and define $S(x)=d(x_{*},x).$\begin{eqnarray*}
\lim_{n\rightarrow\infty}S(\tau^{n}(x)) & = & \lim_{n\rightarrow\infty}d(x_{*},\tau^{n}(x))\\
 & = & \lim_{n\rightarrow\infty}d(\tau^{n}(x_{*}),\tau^{n}(x))\neq d(x_{*},x)=S(x)\quad\forall x\neq x_{*},\end{eqnarray*}
 hence $S(x)$ is a generalized Lyapunov function. By Theorem \ref{thm:weaklyapov}
it follows that $\tau$ is mixing. Secondly, if $\tau$ is mixing,
then \[
\lim_{n\rightarrow\infty}d(\tau^{n}(x),\tau^{n}(y))=d(x_{*},x_{*})=0\neq d(x,y)\quad\forall x,y\in\mathcal{X},\, x\neq y,\]
 so $\tau$ is an asymptotic deformation. \end{proof} \begin{rem*}
Note that the existence of a fixed point is assured if $\tau$ is
a weak contraction on a compact space~\cite{STAKGOLD}, or if the
metric space is convex compact \cite{DUGUNDJI}. As a special case
it follows that any weak contraction $\tau$ on a compact metric space
is mixing. This result can be seen as an instance of Banach contraction
principle on compact spaces. In the second part of the paper we will
present a counterexample which shows that weak contractivity is only
a sufficient criterion for mixing (see Example~\ref{exm:mixing}).
In the context of quantum channels an analogous criterion was suggested
in~\cite{STRICTCONTRATIONS,RAGINSKY} which applied to strict contractions.
We also note that for weak and strict contractions, the trivial generalized
Lyapunov function (Theorem \ref{thm:glyap}) is a strict Lyapunov
function. \end{rem*}

\section{Quantum Channels}

\label{QC}

\begin{figure}[t]
\begin{centering}\includegraphics[width=0.9\columnwidth]{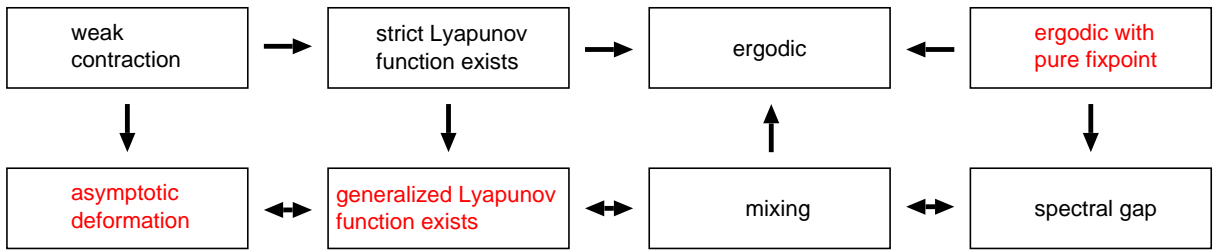}\par\end{centering}

\caption{\label{fig:relations}Relations between the different properties
of a quantum channel. The red text indicates the new results obtained
in this paper, and the black text indicates formerly known results
which we reviewed.}
\end{figure}

In this Section we discuss the mixing properties of quantum channels~\cite{NIELSEN}
which account for the most general evolution a quantum system can
undergo including measurements and coupling with external environments.
In this context solving the mixing problem~(\ref{mixing0}) is equivalent
to determine if repetitive application of a certain physical transformation
will drive any input state of the system (i.e. its density matrices)
into a unique output configuration. The relationship between the different
mixing criteria one can obtain in this case is summarized in Fig.~\ref{fig:relations}.

At a mathematical level quantum channels correspond to linear maps
acting on the density operators $\rho$ of the system and satisfying
the requirement of being completely positive and trace preserving
(CPT). For a formal definition of these properties we refer the reader
to~\cite{KRAUS,WERNER,KEYL}: here we remind only that a necessary
and sufficient conditions to being CPT is to allow Kraus decomposition~\cite{KRAUS}
or, equivalently, Stinespring dilation~\cite{STINE}. Our results
are applicable if the underlying Hilbert space is finite dimensional.
In such regime there is no ambiguity in defining the convergence of
a sequence since all operator norms are equivalent (i.e. given two
norms one can construct an upper and a lower bound for the first one
by properly scaling the second one). Also the set of bounded operators
and the set of operators of Hilbert-Schmidt class coincide. For the
sake of definiteness, however, we will adopt the trace-norm which,
given the linear operator $\Theta:\mathcal{H}\rightarrow\mathcal{H}$,
is defined as $\|\Theta\|_{1}=\mbox{Tr}[\sqrt{\Theta^{\dag}\Theta}]$
with $\mbox{Tr}[\cdots]$ being the trace over $\mathcal{H}$ and
$\Theta^{\dag}$ being the adjoint of $\Theta$. This choice is in
part motivated by the fact~\cite{RUSKAI} that any quantum channel
is non-expansive with respect to the metric induced%
\footnote{This is just the trace distance $d(\rho,\sigma)=\|\rho-\sigma\|_{1}$.%
} by $\|\cdot\|_{1}$ (the same property does not necessarily apply
to other operator norms, e.g. the Hilbert-Schmidt norm, also when
these are equivalent to $\|\cdot\|_{1}$).

We start by showing that the mixing criteria discussed in the first
half of the paper do apply to the case of quantum channels. Then we
will analyze these maps by studying their linear extensions in the
whole vector space formed by the linear operators of $\mathcal{H}$.
Similar questions also arise in the context of finitely correlated
states, where one investigates the decay of correlations in space
(rather than in time)~\cite{FANNES}. 

%


\subsection{Mixing criteria for Quantum Channels}

\label{sec:mixing}

Let $\mathcal{H}$ be a finite dimensional Hilbert space and let $\mathcal{S}(\mathcal{H})$
be the set of its density matrices $\rho$. The latter is a convex
and compact subspace of the larger normed vector space $\mathcal{L}(\mathcal{H})$
composed by the linear operators $\Theta:\mathcal{H}\rightarrow\mathcal{H}$
of $\mathcal{H}$. From this and from the fact that CPT maps are continuous
(indeed they are linear) it follows that for a quantum channel there
always exists at least one density operator which is a fixed point~\cite{TERHAL}.
It also follows that all the results of the previous section apply
to quantum channels. In particular Lemma~\ref{lem:hausdorff} holds
implying that any mixing quantum channel must be ergodic. The following
example shows however that it is possible to have ergodic quantum
channels which are not mixing.

\begin{exm}\label{exm:ergodic} Consider the qubit quantum channel
$\tau$ obtained by cascading a completely decoherent channel with
a NOT gate. Explicitly $\tau$ is defined by the transformations $\tau(|0\rangle\langle0|)=|1\rangle\langle1|$,
$\tau(|1\rangle\langle1|)=|0\rangle\langle0|$, and $\tau(|0\rangle\langle1|)=\tau(|1\rangle\langle0|)=0$
with $|0\rangle,|1\rangle$ being the computational basis of the qubit.
This map is ergodic with fixed point given by the completely mixed
state $(|0\rangle\langle0|+|1\rangle\langle1|)/2$. However it is
trivially not mixing since, for instance, repetitive application of
$\tau$ on $|0\rangle\langle0|$ will oscillate between $|0\rangle\langle0|$
and $|1\rangle\langle1|$. \end{exm}

Theorems~\ref{thm:weakBanach} implies that a quantum channel $\tau:\mathcal{S}(\mathcal{H})\rightarrow\mathcal{S}(\mathcal{H})$
is mixing if and only if it is an asymptotic deformation. As already
pointed out in the introduction, this property is \emph{metric independent}
(as opposed to contractivity). Alternatively, if the fixed point of
a quantum channel is known, then one may use the trivial generalized
Lyapunov function (Theorem~\ref{thm:glyap}) to check if it is mixing.
However both criteria depend on the metric distance, which usually
has no easy physical interpretation. A more useful choice of the is
the quantum relative entropy, which is defined as \begin{eqnarray}
H(\rho,\sigma)\equiv\textrm{Tr}\rho(\log\rho-\log\sigma).\end{eqnarray}
 As discussed in~\cite{JENS}, the quantum relative entropy is continuous
in finite dimension and can be used as a measure of \char`\"{}distance\char`\"{}
(though it is not a metric). It is finite if the support of $\rho$
is contained in the support of $\sigma$. To ensure that it is a continuous
function on a compact space, we consider the case when $\sigma$ is
faithful: \begin{thm}[Relative entropy criterion] A quantum channel
with faithful fixed point $\rho_{*}$ is mixing if and only if the
quantum relative entropy with respect to $\rho_{*}$ is a generalized
Lyapunov function.\end{thm} \begin{proof} Because of Theorem~\ref{thm:weaklyapov}
we only need to prove the second part of the thesis, i.e. that mixing
channels admit the quantum relative entropy with respect to the fixed
point, $S(\rho)\equiv H(\rho,\rho_{*})$, as generalized Lyapunov
function. Firstly notice that the quantum relative entropy is monotonic
under quantum channels~\cite{RUSKAI2}. Therefore the limit $S_{*}(\rho)\equiv\lim_{n\rightarrow\infty}S\left(\tau^{n}(\rho)\right)$
does exist and satisfies the condition $S_{*}(\rho)\geqslant S(\rho)$.
Suppose now there exists a $\rho$ such that $S_{*}(\rho)=S(\rho)$.
Because $\tau$ is mixing and $S$ is continuous we have \[
S(\rho)=S_{*}(\rho)=\lim_{n\rightarrow\infty}S\left(\tau^{n}(\rho)\right)=S(\rho_{*})=0,\]
 and hence $H(\rho,\rho_{*})=0$. Since $H(\rho,\sigma)=0$ if and
only if $\rho=\sigma$ it follows that $S$ is a Lyapunov function
around $\rho_{*}$. \end{proof}\begin{cor}[Unital channels] A unital
channel is mixing if and only if the von Neumann entropy is a generalized
Lyapunov function. \end{cor} Another sufficient condition for mixing
is weak contractivity. As already mentioned in the previous section,
unfortunately this not a necessary condition. Here we present an explicit
counterexample based on a quantum channel introduced in Ref.~\cite{TERHAL}.

\begin{exm}\label{exm:mixing} Consider a three-level quantum system
characterized by the orthogonal vectors $|0\rangle,|1\rangle,|2\rangle$
and the quantum channel $\tau$ defined by the transformations $\tau(|2\rangle\langle2|)=|1\rangle\langle1|$,
$\tau(|1\rangle\langle1|)=\tau(|0\rangle\langle0|)=|0\rangle\langle0|$,
and $\tau(|i\rangle\langle j|)=0$ for all $i\neq j$. It's easy to
verify that after just two iterations any input state $\rho$ will
be transformed into the vector $|0\rangle\langle0|$. Therefore the
map is mixing. On the other hand it is explicitly not a weak contraction
with respect to the trace norm since, for instance, one has \[
\|\;\tau(|2\rangle\langle2|)-\tau(|0\rangle\langle0|)\;\|_{1}=\|\;|1\rangle\langle1|-|0\rangle\langle0|\;\|_{1}=\|\;|2\rangle\langle2|-|0\rangle\langle0|\;\|_{1}\;,\]
 where in the last identity we used the invariance of $\|\cdot\|_{1}$
with respect to unitary transformations. \end{exm}

\subsection{Beyond the density matrix operator space: spectral properties}

\label{sec:newmixing}

Exploiting linearity quantum channels can be extended beyond the space
$\mathcal{S}(\mathcal{H})$ of density operators to become maps defined
on the full vector space $L(\mathcal{H})$ of the linear operators
of the system, in which basic linear algebra results hold. This allows
one to simplify the analysis even though the mixing property~(\ref{mixing0})
is still defined with respect to the density operators of the system.

Mixing conditions for quantum channels can be obtained by considering
the structure of their eigenvectors in the extended space $\mathcal{L}(\mathcal{H})$.
For example, it is easily shown that the spectral radius~\cite{HORNJOHNSON}
of any quantum channel is equal to unity~\cite{TERHAL}, so its eigenvalues
are contained in the unit circle. The eigenvalues $\lambda$ on the
unit circle (i.e. $|\lambda|=1$) are referred to as \emph{peripheral
eigenvalues.} Also, as already mentioned, since $\mathcal{S}(\mathcal{H})$
is compact and convex, CPT maps have always at least one fixed point
which is a density matrix~\cite{TERHAL}. A well-known connection
between the mixing properties and the spectrum is given by the

\begin{thm}[Spectral gap criterion]\label{thm:peri} A quantum channel
is mixing if and only if its only peripheral eigenvalue is $1$ and
this eigenvalue is simple. \end{thm} \begin{proof} The {}``if''
direction can be found in linear algebra textbooks (see for example~\cite[Lemma 8.2.7]{HORNJOHNSON}.
Now let us assume that $\tau$ is a mixing quantum channel with fixed
point $\rho_{*}$. Let $\Theta$ be a generic operator in $\mathcal{L}(\mathcal{H})$.
Then $\Theta$ can be decomposed in a finite set of non-orthogonal
density operators%
\footnote{To show that this is possible, consider an arbitrary operator basis
of $\mathcal{L}(\mathcal{H})$. If $N$ is the finite dimension of
$\mathcal{H}$ the basis will contain $N^{2}$ elements. Each element
of the basis can then be decomposed into two Hermitian operators,
which themselves can be written as linear combinations of at most
$N$ projectors. Therefore there exists a generating set of at most
$2N^{3}$ positive operators, which can be normalized such that they
are quantum states. There even exists a basis (i.e. a minimal generating
set), but in general it can not be orthogonalized.%
}, i.e. $\Theta=\sum_{\ell}c_{\ell}\rho_{\ell}$, with $\rho_{\ell}\in\mathcal{S}(\mathcal{H})$
and $c_{\ell}$ complex. Since $\textrm{Tr}\left[\rho_{\ell}\right]=1$,
we have have $\textrm{Tr}\left[\Theta\right]=\sum_{\ell}c_{\ell}$.
Moreover since $\tau$ is mixing we have $\lim_{n\rightarrow\infty}\tau^{n}\left(\rho_{\ell}\right)=\rho_{*}$
for all $\ell$, with convergence with respect to the trace-norm.
Because of linearity this implies \begin{eqnarray}
\lim_{n\rightarrow\infty}\tau^{n}\left(\Theta\right)=\sum_{\ell}c_{\ell}\;\rho_{*}=\textrm{Tr}\left[\Theta\right]\;\rho_{*}\;.\label{limit}\end{eqnarray}
 If there existed any other eigenvector $\Theta_{*}$ of $\tau$ with
eigenvalue on the unit circle, then $\lim_{n\rightarrow\infty}\tau^{n}(\Theta_{*})$
would not satisfy Eq.~(\ref{limit}). \end{proof}

The speed of convergence can also be estimated by~\cite{TERHAL}
\begin{eqnarray}
\|\tau^{n}\left(\rho\right)-\rho_{*}\|_{1}\;\leqslant C_{N}\; n^{N}\;\kappa^{n}\;,\label{speed}\end{eqnarray}
 where $N$ is the dimensionality of the underlying Hilbert space,
$\kappa$ is the modulus of the second largest eigenvalue of $\tau$,
and $C_{N}$ is some constant depending only on $N$ and on the chosen
norm. Hence, for $n\gg N$ the convergence becomes exponentially fast.
As mentioned in~\cite{RAGINSKY}, the criterion of Theorem~\ref{thm:peri}
is in general difficult to check. This is because one has to find
all eigenvalues of the quantum channel, which is hard especially in
the high dimensional case. Also, if one only wants to check if a particular
channel is mixing or not, then the amount of information obtained
is much higher than the required amount.

\begin{exm} As an application consider the non mixing CPT map of
Example~\ref{exm:ergodic}. One can verify that apart from the eigenvalue
$1$ associated with its fixed point (i.e. the completely mixed state),
it possess another peripheral eigenvalue. This is $\lambda=-1$ which
is associated with the Pauli operator $|0\rangle\langle0|-|1\rangle\langle1|$.
\end{exm}

\begin{cor}\label{cor:speed} The convergence speed of any mixing
quantum channel is exponentially fast for sufficiently high values
of $n$. \end{cor} \begin{proof} From Theorem~\ref{thm:peri} mixing
channels have only one peripheral eigenvalue and it is simple. Therefore
the derivation of Ref.~\cite{TERHAL} applies and Eq.~(\ref{speed})
holds. \end{proof} This result should be compared with the case of
strictly contractive quantum channels whose convergence were shown
to be exponentially fast along to whole trajectory~\cite{RAGINSKY,STRICTCONTRATIONS}.

\subsection{Ergodic channels with pure fixed points}

\label{sec:pure}

An interesting class of ergodic quantum channel is formed by those
CPT maps whose fixed point is a \emph{pure} density matrix. Among
them we find for instance the maps employed in the communication protocols
of Refs.~\cite{MEMORYSWAP,DUALRAIL,RANDOMRAIL,MULTIRAIL} or those
of the purification schemes of Refs.~\cite{YUASA2,YUASA1}. We now
show that within this particular class, ergodicity and mixing are
indeed equivalent properties. \begin{thm}[Purely ergodic maps] \label{thm:ergodicmixing}Let
$|\psi_{1}\rangle\langle\psi_{1}|$ be the pure fixed point of an
ergodic quantum channel $\tau$. It follows that $\tau$ is mixing.
\end{thm} \begin{proof} We will use Theorem \ref{thm:peri} showing
that $|\psi_{1}\rangle\langle\psi_{1}|$ is the only eigenvector of
$\tau$ with peripheral eigenvalue. Assume in fact that $\Theta\in\mathcal{L}(\mathcal{H})$
is a eigenvector of $\tau$ with peripheral eigenvalue, i.e. \begin{eqnarray}
\tau\left(\Theta\right)=e^{i\varphi}\Theta\;.\label{identity1}\end{eqnarray}
 From Lemma~\ref{lem:OBS5} of the Appendix we know that the density
matrix $\rho=\sqrt{\Theta\Theta^{\dag}}/g$, with $g=\mbox{Tr}\left[\sqrt{\Theta^{\dag}\Theta}\right]>0$,
must be a fixed point of $\tau$. Since this is an ergodic map we
must have $\rho=|\psi_{1}\rangle\langle\psi_{1}|$. This implies $\Theta=g|\psi_{1}\rangle\langle\psi_{2}|$,
with $|\psi_{2}\rangle$ some normalized vector of $\mathcal{H}$.
Replacing it into Eq.~(\ref{identity1}) and dividing both terms
by $g$ yields $\tau\left(|\psi_{1}\rangle\langle\psi_{2}|\right)=e^{i\varphi}|\psi_{1}\rangle\langle\psi_{2}|$
and \begin{eqnarray*}
|\langle\psi_{1}|\tau(|\psi_{1}\rangle\langle\psi_{2}|)|\psi_{2}\rangle|=1\;.\end{eqnarray*}
 Introducing a Kraus set $\{ K_{n}\}_{n}$ of $\tau$ and employing
Cauchy-Schwartz inequality one can then write \begin{eqnarray*}
1 & = & |\langle\psi_{1}|\tau(|\psi_{1}\rangle\langle\psi_{2}|)|\psi_{2}\rangle|=|\sum_{n}\langle\psi_{1}|K_{n}|\psi_{1}\rangle\langle\psi_{2}|K_{n}^{\dag}|\psi_{2}\rangle|\\
 & \leqslant & \sqrt{\sum_{n}\langle\psi_{1}|K_{n}|\psi_{1}\rangle\langle\psi_{1}|K_{n}^{\dag}|\psi_{1}\rangle}\sqrt{\sum_{n}\langle\psi_{2}|K_{n}|\psi_{2}\rangle\langle\psi_{2}|K_{n}^{\dag}|\psi_{2}\rangle}\\
 & = & \sqrt{\langle\psi_{1}|\tau(|\psi_{1}\rangle\langle\psi_{1}|)|\psi_{1}\rangle}\sqrt{\langle\psi_{2}|\tau(|\psi_{2}\rangle\langle\psi_{2}|)|\psi_{2}\rangle}=\sqrt{\langle\psi_{2}|\tau(|\psi_{2}\rangle\langle\psi_{2}|)|\psi_{2}\rangle}\;,\end{eqnarray*}
 where we used the fact that $|\psi_{1}\rangle$ is the fixed point
of $\tau$. Since $\tau$ is CPT the quantity $\langle\psi_{2}|\tau(|\psi_{2}\rangle\langle\psi_{2}|)|\psi_{2}\rangle$
is upper bounded by $1$. Therefore in the above expression the inequality
must be replaced by an identity, i.e. \begin{eqnarray*}
\langle\psi_{2}|\tau(|\psi_{2}\rangle\langle\psi_{2}|)|\psi_{2}\rangle=1\qquad\Longleftrightarrow\qquad\tau(|\psi_{2}\rangle\langle\psi_{2}|)=|\psi_{2}\rangle\langle\psi_{2}|\;.\end{eqnarray*}
 Since $\tau$ is ergodic, we must have $|\psi_{2}\rangle\langle\psi_{2}|=|\psi_{1}\rangle\langle\psi_{1}|$.
Therefore $\Theta\propto|\psi_{1}\rangle\langle\psi_{1}|$ which shows
that $|\psi_{1}\rangle\langle\psi_{1}|$ is the only peripheral eigenvalue
of $\tau$. \end{proof}

An application of the previous Theorem is obtained as follows.

\begin{lem}\label{lem:add} Let $M_{AB}=M_{A}\otimes1_{B}+1_{A}\otimes M_{B}$
be an observable of the composite system $\mathcal{H}_{A}\otimes\mathcal{H}_{B}$
and $\tau$ the CPT linear map on $\mathcal{H}_{A}$ of Stinespring
form~\emph{\cite{STINE}} \begin{equation}
\tau(\rho)=\emph{\mbox{Tr}}_{B}\left[U\left(\rho\otimes|\phi\rangle_{B}\langle\phi|\right)U^{\dag}\right]\;,\label{eq:rep}\end{equation}
 (here $\emph{\mbox{Tr}}_{X}\left[\cdots\right]$ is the partial trace
over the system $X$, and $U$ is a unitary operator of $\mathcal{H}_{A}\otimes\mathcal{H}_{B}$).
Assume that $\left[M_{AB},U\right]=0$ and that $|\phi\rangle_{B}$
is the eigenvector corresponding to a non-degenerate maximal or minimal
eigenvalue of $M_{B}.$ Then $\tau$ is mixing if and only if $U$
has one and only one eigenstate that factorizes as $|\nu\rangle_{A}\otimes|\phi\rangle_{B}.$
\end{lem} \begin{proof} Let $\rho$ be an arbitrary fixed point
of $\tau$ (since $\tau$ is CPT it has always at least one), i.e.
$\textrm{Tr}_{B}\left[U\left(\rho\otimes|\phi\rangle_{B}\langle\phi|\right)U^{\dag}\right]=\rho$.
Since $M_{AB}=M_{A}+M_{B}$ is conserved, and $\textrm{Tr}_{A}\left[M_{A}\rho\right]=\textrm{Tr}_{A}\left[M_{A}\tau(\rho)\right]$,
the expectation value of $M_{B}$ is unchanged. Hence system $B$
must remain in the state with maximal/minimal eigenvalue, which we
have assumed to be unique and pure, i.e. \[
U\left(\rho\otimes|\phi\rangle_{B}\langle\phi|\right)U^{\dag}=\rho\otimes|\phi\rangle_{B}\langle\phi|\qquad\Longrightarrow\qquad\left[U,\rho\otimes|\phi\rangle_{B}\langle\phi|\right]=0\;.\]
 Thus there exists a orthonormal basis $\left\{ |u_{k}\rangle\right\} _{k}$
of $\mathcal{H}_{A}\otimes\mathcal{H}_{B}$ diagonalizing simultaneously
both $U$ and $\rho\otimes|\phi\rangle_{B}\langle\phi|$. We express
the latter in this basis, i.e. $\rho\otimes|\phi\rangle_{B}\langle\phi|=\sum_{k}p_{k}|u_{k}\rangle\langle u_{k}|,$
and perform the partial trace over subsystem $A$ to get\[
|\phi\rangle_{B}\langle\phi|=\sum_{k}p_{k}\mbox{Tr}_{A}\left[|u_{k}\rangle\langle u_{k}|\right].\]
Hence $\textrm{Tr}_{A}\left[|u_{k}\rangle\langle u_{k}|\right]=|\phi\rangle_{B}\langle\phi|$
for all $k,$ and $|u_{k}\rangle$ must be factorizing, \begin{equation}
|u_{k}\rangle=|\nu_{k}\rangle_{A}\otimes|\phi\rangle_{B}.\label{eq:fact}\end{equation}
 If the factorizing eigenstate of $U$ is unique, it follows that
$\rho=|\nu\rangle\langle\nu|$ for some $|\nu\rangle$ and that $\tau$
is ergodic. By Theorem \ref{thm:ergodicmixing} it then follows that
$\tau$ is also mixing. If on the other hand there exists more than
one factorizing eigenstate, then all states of the form of Eq. (\ref{eq:fact})
correspond to a fixed point $\rho_{k}=|\nu_{k}\rangle\langle\nu_{k}|$
and $\tau$ is neither ergodic nor mixing. \end{proof}

The case discussed in Lemma~\ref{lem:add} is a generalization of
the CPT map discussed in Ref.~\cite{MEMORYSWAP} in the context of
spin-chain communication. There $\mathcal{H}_{A}$ and $\mathcal{H}_{B}$
represented two distinct part of a chain of spins coupled through
Heisenberg-like interactions: the latter including the spins controlled
by the receiver of the message, while the former accounting for all
the remaining spins. Assuming the system to be initially in the ground
state (i.e. all spin down), the sender (located at one of the extremes
of the chain) encodes her/his quantum messages (i.e. qubits) into
superpositions of spins excitations which will start propagating toward
to receiver (located at the other extreme of the chain). In Ref.~\cite{MEMORYSWAP}
it was shown that, by repetitively swapping the spins which are under
her/his control with some ancillary spins prepared in the ground state,
the receiver will be able to recover the transmitted messages. The
key ingredient of such result is the fact that by applying the swapping
operations the receiver is indeed removing all the excitations (and
therefore the corresponding encoded quantum information) out that
part of the chain which is not directly accessible to him/her (i.e.
the part represented by $\mathcal{H}_{A}$). In its simplest version,
the resulting transformation on $\mathcal{H}_{A}$ can be described
by Eq.~(\ref{eq:rep}) with $U$ and $M_{AB}$ representing, respectively,
the free evolution of the spins among two consecutive swaps and the
z-component of the magnetization of the chain. Lemma~\ref{lem:add}
can then be used to provide an alternative proof of convergence of
the protocol~\cite{MEMORYSWAP} showing that indeed repetitive applications
of $\tau$ will drive $\mathcal{H}_{A}$ toward a unique convergence
point (i.e. the state with no excitation).

\section{Conclusion}

\label{CONCLUSION} In reviewing some known results on the mixing
property of continuous maps, we derived a stronger version of the
direct Lyapunov method. For compact metric spaces (including quantum
channels operating over density matrices) it provides a necessary
and sufficient condition for mixing. Moreover it allows us to prove
that asymptotic deformations with at least one fixed point must be
mixing.

In the specific context of quantum channels we employed the generalized
Lyapunov method to analyze the mixing properties. Here we also analyzed
different mixing criteria. In particular we have shown that an ergodic
quantum channels with a pure fixed point is also mixing.

\begin{acknowledgement*} DB acknowledges the support of the UK Engineering
and Physical Sciences Research Council, Grant Nr. GR/S62796/01 and
Scuola Normale Superiore Grant {}``Giovani Ricercatori (2005/2006)''.
He would like to thank Sougato Bose and Heinz-Peter Breuer for stimulating
discussions and Vittorio Giovannetti for his kind hospitality. VG
contribution was in part supported by the Quantum Information research
program of Centro di Ricerca Matematica Ennio De Giorgi of Scuola
Normale Superiore.

\end{acknowledgement*}

\appendix

\section*{Appendix}

Here we derive some Lemmas which are not correlated with each other
but which are relevant in our discussion. Lemma~\ref{lem:subsequences}
discusses a property of sequentially compact topological spaces. Lemma~\ref{thm:ergodic}
states a well known theorem \cite{RAGINSKY} which, in the context
of normed vector spaces, shows the equivalence between the definition
of ergodicity of Eq.~(\ref{defergo10}) and its definition using
time averages. Finally Lemma~\ref{lem:OBS5} discusses a useful property
of quantum channels (see also \cite{SCHRADER}).

\begin{lem}\label{lem:subsequences} Let $x_{n}$ be a sequence in
a sequentially compact topological space $\mathcal{X}$ such that
any convergent subsequence converges to $x_{*}.$ Then the sequence
converges to $x_{*}.$ \end{lem} \begin{proof} We prove by contradiction:
assume that the sequence does not converge to $x_{*}.$ Then there
exists an open neighborhood $O(x_{*})$ of $x_{*}$ such that for
all $k\in\mathbb{N},$ there is a $n_{k}$ such that $x_{n_{k}}\notin O(x_{*}).$
Thus the subsequence $x_{n_{k}}$ is in the closed space $\mathcal{X}\backslash O(x_{*}),$
which is again sequentially compact. $x_{n_{k}}$ has a convergent
subsequence with a limit in $\mathcal{X}\backslash O(x_{*}),$ in
particular this limit is not equal to $x_{*}.$ \end{proof}

\begin{lem} \label{thm:ergodic}Let $\mathcal{X}$ be a convex compact
subset of a normed vector space, and let $\tau:\mathcal{X}\rightarrow\mathcal{X}$
be a continuous map. If $\tau$ is ergodic with fixed point $x_{*},$
then \begin{eqnarray}
\lim_{n\rightarrow\infty}\frac{1}{n+1}\sum_{\ell=0}^{n}\tau^{\ell}(x)=x_{*}\;.\label{average}\end{eqnarray}
 \end{lem} \begin{proof} Define the sequence $A_{n}\equiv\frac{1}{n+1}\sum_{\ell=0}^{n}\tau^{\ell}(x)$.
Let then $M$ be the upper bound for the norm of vectors in $\mathcal{X}$,
i.e. $M\equiv\sup_{x\in\mathcal{X}}\| x\|<\infty$. which exists because
$\mathcal{X}$ is compact. The sequence $A_{n}$ has a convergent
subsequence $A_{n_{k}}$ with limit $\tilde{A}.$ Since $\tau$ is
continuous one has $\lim_{k\rightarrow\infty}\tau(A_{n_{k}})=\tau(\tilde{A})$.
On the other hand, we have\[
\|\tau(A_{n_{k}})-A_{n_{k}}\|=\frac{1}{n_{k}+1}\|\tau^{n_{k}+1}(x)-x\|\leqslant\frac{\|\tau^{n_{k}+1}(x)\|+\| x\|}{n_{k}+1}\leqslant\frac{2M}{n_{k}+1},\]
 so the two sequences must have the same limit, i.e. $\tau(\tilde{A})=\tilde{A}$.
Since $\tau$ is ergodic, we have $\tilde{A}=x_{*}$ and $\lim_{n\rightarrow\infty}A_{n}=x_{*}$
by Lemma \ref{lem:subsequences}. \end{proof} \begin{rem*} Note
that if $\tau$ has a second fixed point $y_{*}\neq x_{*}$, then
for all $n$ one has $\frac{1}{n+1}\sum_{\ell=0}^{n}\tau^{\ell}(y_{*})=y_{*}$,
so Eq.~(\ref{average}) would not apply. \end{rem*}

\begin{lem}\label{lem:OBS5} Let $\tau$ be a quantum channel and
$\Theta$ be an eigenvector of $\tau$ with peripheral eigenvalue
$\lambda=e^{i\varphi}$. Then, given $g=\mbox{\emph{Tr}}\left[\sqrt{\Theta^{\dag}\Theta}\right]>0$,
the density matrices $\rho=\sqrt{\Theta\Theta^{\dag}}/g$ and $\sigma=\sqrt{\Theta^{\dag}\Theta}/g$
are fixed points of $\tau$. \end{lem} \begin{proof} Use the left
polar decomposition to write $\Theta=g\;\rho U$ where $U$ is a unitary
operator. The operator $\rho U$ is clearly an eigenvector of $\tau$
with eigenvalue $e^{i\varphi}$, i.e. \begin{eqnarray}
\tau(\rho U)=\lambda\;\rho U\;.\label{uuu}\end{eqnarray}
 Hence introducing a Kraus set $\{ K_{n}\}_{n}$ of $\tau$~\cite{KRAUS}
and the spectral decomposition of the density matrix $\rho=\sum_{j}p_{j}|\psi_{j}\rangle\langle\psi_{j}|$
with $p_{j}>0$ being its positive eigenvalues, one gets \begin{eqnarray*}
\lambda=\mbox{Tr}[\tau(\rho U)U^{\dag}]=\sum_{j,\ell,n}p_{j}\langle\phi_{\ell}|K_{n}|\psi_{j}\rangle\langle\psi_{j}|UK_{n}^{\dag}U^{\dag}|\phi_{\ell}\rangle\;,\end{eqnarray*}
 where the trace has been performed with respect to an orthonormal
basis $\{|\phi_{\ell}\rangle\}_{\ell}$ of $\mathcal{H}$. Taking
the absolute values of both terms gives \begin{eqnarray}
|\lambda| & = & |\sum_{j,\ell,n}p_{j}\langle\phi_{\ell}|K_{n}|\psi_{j}\rangle\langle\psi_{j}|UK_{n}^{\dag}U^{\dag}|\phi_{\ell}\rangle|\nonumber \\
 & \leqslant & \sqrt{\sum_{j,\ell,n}p_{j}\langle\phi_{\ell}|K_{n}|\psi_{j}\rangle\langle\psi_{j}|K_{n}^{\dag}|\phi_{\ell}\rangle}\sqrt{\sum_{j,\ell,n}p_{j}\langle\phi_{\ell}|UK_{n}U^{\dag}|\psi_{j}\rangle\langle\psi_{j}|UK_{n}^{\dag}U^{\dag}|\phi_{\ell}\rangle}\label{cauchysw}\\
 & = & \sqrt{\mbox{Tr}[\tau(\rho)}]\sqrt{\mbox{Tr}[\tilde{\tau}(\rho)]}=1,\nonumber \end{eqnarray}
 where the inequality follows from the Cauchy-Schwartz inequality.
The last identity instead is a consequence of the fact that the transformation
$\tilde{\tau}(\rho)=U\tau(U^{\dag}\rho U)U^{\dag}$ is CPT and thus
trace preserving. Since $|\lambda|=1$ it follows that the inequality~(\ref{cauchysw})
must be replaced by an identity. This happens if and only if there
exist $e^{i\vartheta}$ such that \begin{eqnarray*}
\sqrt{p_{j}}\{\langle\phi_{\ell}|K_{n}|\psi_{j}\rangle\}^{*}=\sqrt{p_{j}}\langle\psi_{j}|K_{n}^{\dag}|\phi_{\ell}\rangle=e^{i\vartheta}\sqrt{p_{j}}\langle\psi_{j}|UK_{n}^{\dag}U^{\dag}|\phi_{\ell}\rangle\;,\end{eqnarray*}
 for all $j,\ell$ and $n$. Since the $|\phi_{\ell}\rangle$ form
a basis of $\mathcal{H}$, and $p_{j}>0$ this implies \begin{eqnarray*}
\langle\psi_{j}|K_{n}^{\dag}=e^{i\vartheta}\;\langle\psi_{j}|UK_{n}^{\dag}U^{\dag}\quad\Rightarrow\quad\langle\psi_{j}|UK_{n}^{\dag}=e^{-i\vartheta}\;\langle\psi_{j}|K_{n}^{\dag}U\;,\end{eqnarray*}
 for all $n$ and for all the not null eigenvectors $|\psi_{j}\rangle$
of $\rho$. This yields \begin{eqnarray*}
\tau(\rho U) & = & \sum_{j}p_{j}\sum_{n}K_{n}|\psi_{j}\rangle\langle\psi_{j}|UK_{n}^{\dag}=e^{-i\vartheta}\;\sum_{j}p_{j}\sum_{n}K_{n}|\psi_{j}\rangle\langle\psi_{j}|K_{n}^{\dag}U\\
 & = & e^{-i\vartheta}\;\tau(\rho)U\end{eqnarray*}
 which, replaced in (\ref{uuu}) gives $e^{-i\vartheta}\;\tau(\rho)=e^{i\varphi}\;\rho$,
whose only solution is $e^{-i\vartheta}=e^{i\varphi}$. Therefore
$\tau(\rho)=\rho$ and $\rho$ is a fixed point of $\tau$. The proof
for $\sigma$ goes along similar lines: simply consider the right
polar decomposition of $\Theta$ instead of the left polar decomposition.
\end{proof}

\begin{cor} Let $\tau$ be an ergodic quantum channel. It follows
that its eigenvectors associated with peripheral eigenvalues are normal
operators. \end{cor} \begin{proof} Let $\Theta$ be an eigenoperator
with peripheral eigenvalue $e^{i\varphi}$ such that $\tau\left(\Theta\right)=e^{i\varphi}\;\Theta$.
By Lemma \ref{lem:OBS5} we know that, given $g=\mbox{Tr}\left[\sqrt{\Theta^{\dag}\Theta}\right]$
the density matrices $\rho=\sqrt{\Theta\Theta^{\dag}}/g$ and $\sigma=\sqrt{\Theta^{\dag}\Theta}/g$
must be fixed points of $\tau$. Since the map is ergodic we must
have $\rho=\sigma$, i.e. $\Theta\Theta^{\dag}=\Theta^{\dag}\Theta$.
\end{proof}

\end{document}